# Passive radiative cooling impact on commercial crystalline silicon-based photovoltaics

George Perrakis[*,1,2], Anna C. Tasolamprou[1], George Kenanakis[1], Eleftherios N. Economou[1,3], Stelios Tzortzakis[1,2,4], Maria Kafesaki[1,2]

[1]Institute of Electronic Structure and Laser (IESL), Foundation for Research and Technology-Hellas (FORTH), Heraklion, Greece

[2]Dept. of Materials Science and Technology, Univ. of Crete, Heraklion, Greece

[3]Department of Physics, University of Crete GR-71003, Heraklion, Crete, Greece

[4]Science Program, Texas A&M University at Qatar, P.O. Box 23874, Doha, Qatar

[*]*E-mail: gperrakis@iesl.forth.gr*

**Abstract**

The radiative cooling of objects during daytime under direct sunlight has recently been shown to be significantly enhanced by utilizing nanophotonic coatings. Multilayer thin film stacks, 2D photonic crystals, etc. as coating structures improved the thermal emission rate of a device in the infrared atmospheric transparency window reducing considerably devices' temperature. Due to the increased heating in photovoltaic (PV) devices, that has significant adverse consequences on both their efficiency and life-time, and inspired by the recent advances in daytime radiative cooling, we developed a coupled thermal-electrical modeling to examine the physical mechanisms on how a radiative cooler affects the overall efficiency of commercial photovoltaic modules. Employing this modeling, which takes into account all the major processes affected by the temperature variation in a PV device, we evaluated the relative impact of the main radiative cooling approaches proposed so far on the PV efficiency, and we established required conditions for optimized radiative cooling. Moreover, we identified the validity regimes of the currently existing PV-cooling models which treat the PV coolers as simple thermal emitters. Finally, we assessed some realistic photonic coolers from the literature, compatible with photovoltaics, to implement the radiative cooling requirements, and demonstrated their associated impact on the temperature reduction and PV efficiency. Providing the physical mechanisms and requirements for cooling radiatively solar cells, our study provides guidelines for utilizing suitable photonic structures as radiative coolers, enhancing the efficiency and the lifetime of PV devices.

## 1. Introduction

A solar cell operating under the sun inevitably generates heat apart from electrical power. Principally, the highest fraction of the absorbed sunlight remains unexploited as it was explained by Shockley and Queisser in their seminal paper[1] in 1961. According to their analysis, a single-junction silicon-based (semiconductor material with a band-gap of ~1.1 eV) solar cell has a theoretical upper (SQ) limit for incident solar to electrical power conversion efficiency of around 32%, assuming that it operates at a constant temperature equal to 300 K. In practice, residual power dissipates mainly into heat that increases the operating temperature of the solar cell, leading to substantial adverse consequences not only for the lifetime of the materials, but also for the efficiency of the system mainly due to the increased carrier recombination at elevated temperatures[2]. The heating problem becomes even more prominent in conventional photovoltaic systems (PVs) due to the accumulated heat that arises from the parasitic absorption of incident photons at the various parts of the PV device (see Fig. 1a) other than the semiconductor material. This occurs not only at the wavelengths within the absorption band of the semiconductor (for silicon: ~0.28-1.1 μm) but also beyond these wavelengths (sub-bandgap radiation, ~1.1-4 μm, which is a heat source) where the sun still has considerable intensity. As a result, typical operating temperatures[3] can reach values even higher than ~325 K. Indicatively, for a crystalline silicon solar cell, every 1 K temperature rise leads to a relative efficiency decline[4] of about 0.45%. Moreover, the aging rate of a solar cell array doubles for every 10K solar cell temperature increase[5].

The significant adverse consequences of the temperature rise on the solar cells have led to the utilization of several cooling approaches. Conventional strategies for cooling are mainly focused on nonradiative heat transfer via conduction or convection, like forced air flow[6], water cooling[7], heat-pipe-based systems[8], etc., most of which consume extra energy. Recently though, there has been a significant advance in the field of passive (i.e., no extra energy input needed) radiative cooling, targeting though mainly cooling of buildings. Raman et al.[9] in 2014 developed a passive radiative cooling system based on a photonic crystal. The photonic crystal was designed to reflect the solar heating power (~0.28-4 μm) and at the same time allow radiative cooling through thermal emission in the mid-IR, at the atmospheric transparency window of 8-13 μm. In this way the system had radiative access to the coldness of the universe, through this atmospheric transparency window, and therefore could additionally use the universe as a heat sink, with much lower temperature (~3 K) than that of the atmosphere (~300 K). With this approach Raman et al.[9] demonstrated an impressive cooling, up to 5 K under direct sunlight. Subsequently, appropriate passive radiative coolers were designed compatible with PV systems, mainly Si-based PVs, that allowed radiative cooling through enhanced thermal emission in the mid-IR[10,11,12] but at the same time increased the visible light to go through and reach the PV cells[13,14] for only a certain beneficial spectral window (~0.375-1.1 μm), i.e. via enhancing the transmission in this spectral window while reflecting detrimental UV (~0.28-0.375 μm) and sub-bandgap parasitic (~1.1-4 μm) absorption. In this way, the perspective of PV-coolers with a double-role has been demonstrated; both increasing the solar cell absorptivity and also reducing the operating temperature of the device up to ~5.7 K.[13]

As highlighted in Ref.[13], there are currently two major photonic cooling approaches for the radiative thermal management of PVs, focusing on controlling either (i) the solar absorption by reflecting parasitic UV, sub-bandgap radiation and further enhancing the beneficial optical absorption, or (ii) the thermal radiation. Most of the existing studies employing these approaches though treat PVs as solar absorbers and not as quantum devices, i.e. they do not consider the generation of electrical power by the PV, neither all the major temperature-dependent recombination mechanisms of the generated carriers. This, depending on the operation conditions, may lead to an overestimation of the efficiency increase related to the temperature reduction. In our work, we propose a theoretical thermal-electrical co-model, which takes into account all the major processes affected by the temperature variation in a PV device, to examine how a passive radiative cooler affects the overall efficiency of a PV system.

In particular, we analyze and demonstrate the physical mechanisms of the efficiency enhancement related to the temperature reduction in commercial PVs. In this respect, we further distinguish and evaluate the impact of each of the previously mentioned radiative cooling approaches, if implemented both separately and together, on the efficiency enhancement of a PV operating outdoors. Exploiting our model, and particularly its potential to give the impact on the PV efficiency of any different part of the electromagnetic spectrum, we reveal, among others, the currently unexplored considerable impact of the thermalization losses (i.e., excess energy of incident photons relative to the bandgap of the semiconductor that cannot be exploited and finally dissipates into heat) on the PV efficiency. Finally, we examine some realistic photonic structures proposed in the literature towards the implementation and fulfillment of the radiative cooling requirements, and we analyze their impact on the PV efficiency, in comparison also with our evaluated "ideal" cooler.

## 2. Features of solar cell operation in outdoor conditions

In the present work we study the crystalline silicon-based solar cells which are currently dominant in the market of solar cell technology[15]. A typical state-of-the-art silicon-based photovoltaic module along with each interlayer is shown in Fig. 1a. The most important part of the PV module is the cell, where the conversion of the incident solar power to electricity takes place. We assume that the cell involves a 250 μm thick mono-crystalline silicon wafer with interdigitated state-of-the-art type back contacts (IBC) responsible for collecting the photo-generated carriers[16]. All remaining layers, other than the cell, are required for its stable operation. More specifically, the transparent top surface, most often a 3.2 mm thick glass (contains 70−80% silica), protects the exposed solar cell system from the outside conditions and provides mechanical strength and rigidity. The most common encapsulant, the EVA (ethylene-vinyl acetate), is used as a 0.46 mm thick joint that provides adhesion between the cells, the top (glass) and the rear (substrate: made of a 0.5 mm thick Tedlar: polyvinyl fluoride) rough surfaces of the PV module. The main requirements of both the glass and the encapsulant are stability at elevated temperatures and high UV exposure, low thermal resistivity and optical transparency for the incident radiation to reach the cell.

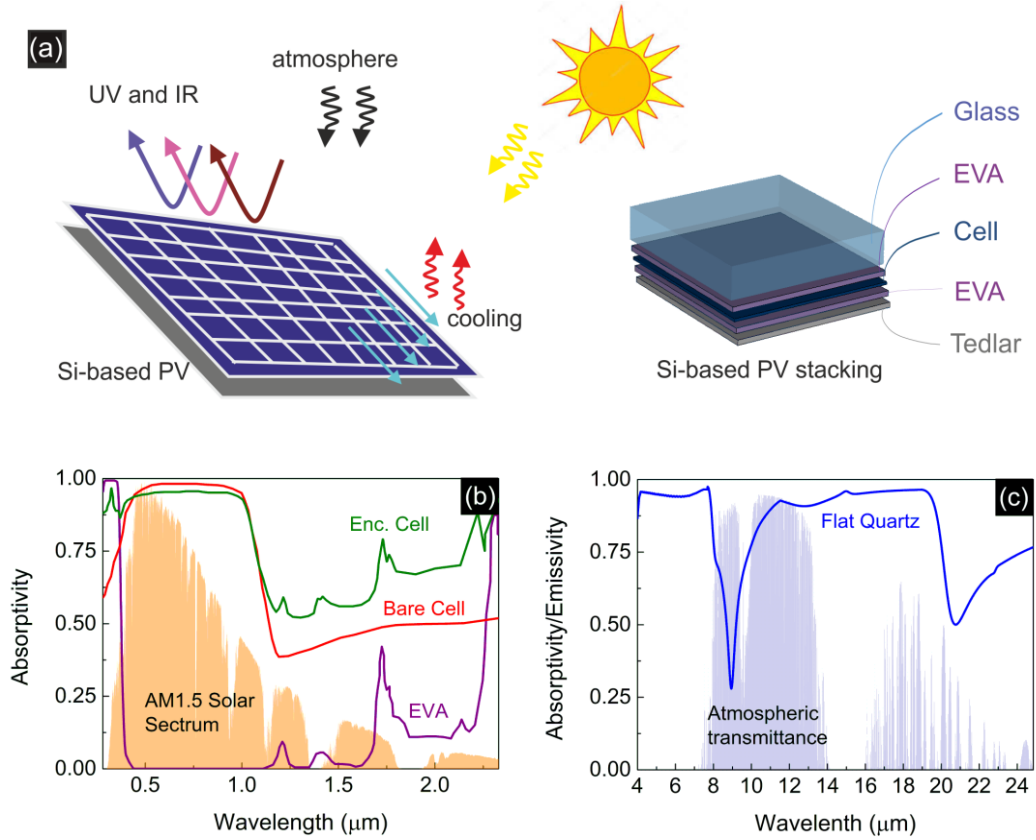

*Fig. 1: (a) Schematic of the cooling approaches for the radiative thermal management of PVs and material stacking of the encapsulated crystalline silicon-based PV. The thickness and role of the different layers of the PV module are discussed in the main text. (b) Absorptivity of the bare cell (red line), encapsulated cell (green line), and of a 0.46 mm thick EVA wafer (purple line). Data are extracted from Refs[13,17,18]. (c) Emissivity spectra in the thermal wavelengths (mid-IR) of a 3.2 mm thick glass (fused Quartz) layer, like the one in panel (a).*

Despite the high optical transparency of both glass and EVA, inevitably the solar absorption spectrum of the encapsulated solar cell (structure shown in Fig. 1a) changes relative to the bare cell, leading to unwanted absorption losses, as shown in Fig. 1b, where the absorption for a bare cell (red line), an encapsulated cell[13] (green line) and a single 0.46 mm EVA wafer[17] (purple line) is shown. It is clear that EVA strongly absorbs UV radiation (~< 0.375 μm) reducing thus the available photo-carriers reaching the cell at this regime, while for the wavelength range within the absorption band of silicon (indirect bandgap of ~1.107 μm) the absorption is slightly reduced mainly due to the reflection (~0.04) introduced from the top surface of the glass. Moreover, unexploitable sub-bandgap absorption, beyond 1.107 μm, up to 4 μm, is still very high, for both the bare and the encapsulated cell, despite that intrinsic silicon does not absorb in this regime. The reason is the non-zero absorption from the highly doped silicon, the metal contacts, the EVA and the thin antireflection layers (usually made of SiN or $SiO_2$) usually placed on top of silicon, together with the light-trapping effect[10,13]. Consequently, sub-bandgap and UV radiation (of intensity ~150 W/m$^2$ according to our simulations) not only remains unexploited but also dissipates into heat, which further reduces the efficiency of the solar cell.

In the mid-IR/thermal wavelength range (4-33 μm), the emissivity spectrum of the encapsulated solar cell is mainly determined by the 3.2 mm thick top glass layer. This emissivity is shown in Fig. 1c, where we have assumed that the glass is fused quartz with permittivity data as given by Palik[18]. Thus, the top protective glass is considered herein as the conventional thermal radiative cooler of the system. The glass exhibits strong phonon-polariton resonances at ~9 μm and ~21 μm that allow it to achieve relatively strong absorptivity/emissivity in the thermal wavelength range of 7-27 μm. On the other hand, nearby the wavelengths of the phonon-polariton resonances (in the ranges of 8-13 μm and 19-30 μm) there is a strong impedance mismatch between glass and air leading to large reflectivity, associated with dips in absorptivity/emissivity. These emissivity dips coincide with the transparency window of the atmosphere (see Fig. 1c), and as a result they lead to reduction of the cooling capability of the system. Therefore, eliminating them is of high importance and has been extensively studied nowadays.

## 3. Electrical - thermal modeling

Crystalline silicon-based solar cells are basically p-n-homojunction diodes, that is a junction of a n-type and p-type doped silicon which possess an excess of free electrons and holes in their carrier concentrations respectively. The forces acting on the electron and hole carriers to produce an electric current are the gradients introduced by the quasi-Fermi energy level splitting ($qV$) in both the n- and p-type material[19] under steady-state non-equilibrium illuminated conditions, since the free electron ($n$) and hole ($p$) carrier concentrations strongly depend upon illumination. Detailed balance method described by Shockley and Queisser[1] relates the current density, $J$ [in A/m$^2$], in ideal solar cells to the output voltage, $V$ [in V], by balancing the particles entering and exiting the device. To this extent, the limiting efficiency of such solar cells is due to the balancing of the number of photons absorbed by the solar cell with the number of carriers exiting the cell either to produce electrical power or to result to emission through radiative recombination of electron-hole pairs. In the present work, besides radiative recombination we further take into consideration the only fundamental nonradiative loss mechanism in mono-crystalline silicon (since for mono-crystalline silicon we assume that there are no defects in the crystalline structure), the Auger recombination. Following Shockley's and Queisser's detailed balance method, the current density obtained in an electrically homogeneous mono-crystalline silicon-based solar cell under illumination can be calculated by

$$J(V,T) = J_0(T)\left(e^{\frac{qV}{k_BT}} - 1\right) + J_A(V,T) - J_{SC} \qquad (1)$$

where $q$ is the elementary charge of an electron [in C], $k_B$ is Boltzmann's constant [in eV/K], $T$ is the operating temperature [in K] and $J_A$ is the nonradiative recombination current density due to Auger recombination. The term

$$J_{SC} = q\int_0^\infty a_{Si}(\lambda)\Phi_{AM1.5G}(\lambda)d\lambda \qquad (2)$$

is the current density flowing at short-circuit conditions under the illumination of the sun. $\Phi_{AM1.5G}$ is the photon flux density [in photons·m$^{-2}$·s$^{-1}$·nm$^{-1}$] of the "AM 1.5G" standard sunlight spectrum[20] reaching the Earth's surface, which is universal when characterizing solar cells. This term equals the photocurrent since in Equation (2) the external quantum efficiency (EQE) of the solar cell (i.e., number of charge carriers collected versus the number of incident photons) is replaced by its absorptivity, $a_{Si}$, owing to the near-unity quantum yield in mono-crystalline silicon-based solar cells[21]. The first term in Equation (1) represents the voltage-dependent radiative recombination current density in the dark. It is a product of the energy distribution of carriers, at a specific operating temperature of the solar cell, that have enough energy to flow through the junction, in the opposite direction from the photogenerated current, and recombine[19]. The energy distribution of carriers and consequently the dark current density follow the Fermi statistics, which, if the Fermi level is lying within the band gap (as in our case), corresponds to Maxwell-Boltzmann statistics. The term $qV$ characterizes the quasi-Fermi energy level splitting, i.e. the difference in the quasi-Fermi levels of electrons and holes (the term "quasi" is due to the non-equilibrium (i.e. under-solar-illumination) steady state). Lastly, $J_0$ is the saturation radiative current density which is independent of bias and it is determined by the thermal excitation level of carriers quantified by the temperature-dependent blackbody (BB) spectrum ($\Phi_{BB}$, see Equation S1a):

$$J_0(T) = q \int a_{Si}(\lambda) \Phi_{BB}(T,\lambda) d\lambda \qquad (3)$$

The Auger recombination rate, which is specific to the chosen semiconductor material, under Boltzmann's approximation and assuming that $n = p$ and $np >> n_i^2$, is given by[22,23]

$$J_A(V,T) = q \cdot 2A_r(T) \cdot n_i^3(T) \cdot e^{\left(\frac{3qV}{2k_BT}\right)} \cdot W \qquad (4)$$

where $W$ is the thickness of the silicon layer, $A_r(T)$ is the temperature-dependent Auger coefficient[24], and $n_i(T)$ is the temperature-dependent intrinsic carrier concentration[25]. Equation (1) assumes that the dark current density remains the same during illuminated conditions and the net current density is shifted in negative current direction by the photocurrent $J_{SC}$ (that flows in the opposite direction from the dark current) which at least for silicon-based solar cells is independent of the bias[26]. In such cases, the superposition (see Equation (1)) between the dark and illuminated $JV$ characteristics of a diode is valid. The efficiency, $\eta$, of the solar cell is given by

$$\eta = \frac{P_{ele,max}}{P_{inc}} = \frac{J_{SC}V_{OC}FF}{P_{inc}} = \frac{J_{mp}V_{mp}}{P_{inc}} \qquad (5)$$

where $P_{ele,max} = \max(-JV) = J_{mp}V_{mp}$ is the electrical power density output of a solar cell operating at the maximum power point[27], $P_{inc}$ is the incident power density of the incoming sun radiation and $FF = J_{mp}V_{mp} / J_{SC}V_{OC}$ is the fill factor. The term $V_{OC}$ is the maximum voltage, usually referred as the open-circuit voltage, and results from Equation (1) by setting the total current $J = 0$ and solving for $V$.

As discussed above, the limiting efficiency of a solar cell depends upon balancing of particles entering and exiting the device for a specific operating temperature of the system. More specifically, assuming only radiative recombination, if the cell operates at high temperature (thus the current $J_0$ becomes higher) the quasi-Fermi energy level splitting ($qV$) must be reduced to maintain a balance between the number of absorbed photons and the number of the emitted photons[28]. This results to lower $V_{OC}$ and $V_{mp}$ and lower efficiencies. Regarding the nonradiative (Auger) recombination process, $J_A$ scales with the intrinsic carrier concentration cubed (see Equation (4)). Therefore, at elevated temperatures the Auger recombination rate is higher due to the increased thermally generated carrier concentrations.

It is important to note here that two of the main assumptions of our theoretical modeling are that we neglect the effect of how efficiently the contacts collect the photo-generated carriers and the impact of the PV defects. Such assumptions are quite valid for calculating the absolute efficiency of mono-crystalline silicon-based solar cells since their internal quantum efficiency (IQE) (i.e., number of charge carriers collected versus the number of incident photons absorbed) is near-unity[21]. Primarily, since we are mostly interested in the efficiency changes owing to the operating temperature variations, studies have shown that the decrease of $\eta$ with temperature is mainly controlled by the reduction of $V_{OC}$ with $T$[29]. The temperature impact on both the contact resistance, PV defects and $J_{SC}$ is negligible, thus leaving the efficiency dependence with temperature to be mainly controlled by the material properties of the semiconductor of the solar cell that have been taken into account in the present study.

To take into consideration the effect of heating in solar cells, and thus to be able to calculate the extracted electrical power or efficiency in respect to the operating temperature at typical outdoor conditions, we perform a thermal analysis. The steady-state temperature or the operating temperature of the cell of a photovoltaic module can be accurately described by treating the PV as a uniform device by using appropriately combined conduction-convection heat transfer coefficients. A thermal analysis for the PV can thus be performed by balancing the total power into and out of the device following Planck's blackbody formalism and Kirchhoff's law, i.e., absorptivity equals emissivity. This strategy (PRC: passive radiative cooling strategy) was firstly proposed by Fan[9,30] for calculating the radiative cooling of solar absorbers and has been shown to exhibit highly accurate results[10]. According to Fan, when a structure is exposed to a daylight sky, it is subject to both solar irradiance and atmospheric thermal radiation (corresponding to ambient air temperature $T_{amb}$). In our case (structure of Fig. 1a), the net cooling power, $P_{net,cool}$, of a PV can be determined by summing the total power into and out of the device[28]:

$$P_{net,cool}(V,T) = P_{rad,cooler}(T) - P_{atm}(T_{amb}) + P_{cond+conv}(T) - P_{solar,heat}(V,T) \quad (6)$$

where $P_{rad,cooler}$ is the power density radiated by the radiative cooler (see Equation S2), i.e. the top glass layer, and $P_{atm}$ is the power density absorbed by the cooler from the atmospheric emission that takes into consideration the atmospheric transparency window[31] (see Equation S3). $P_{cond,conv} = h_c(T - T_{amb})$ is the power density loss (since in our case $T>T_{amb}$) due to convection and conduction, where $h_c = h_{cond} + h_{conv}$ is a combined nonradiative heat transfer coefficient that captures the collective effect of conductive and convective heating owing to the contact of the cell with external surfaces and the air adjacent to the top radiative cooler. The last term, $P_{solar,heat}$, is the absorbed solar power density that dissipates into heat which incorporates the electrical part and formulates as follows:

$$P_{solar,heat}(V,T) = P_{sun} - P_{ele,max}(V,T) - P_{rad,cell}(V,T) \quad (7)$$

In Equation (7) $P_{sun}$ is the total solar absorption power density (see Equation S4), and $P_{rad,cell}$ is the power density radiated by the solar cell also known as the non-thermal radiation (emitted through electron-hole recombination, as a consequence of the bandgap of the semiconductor material[32] - see Equation S5). Consequently, we notice that both the quasi-Fermi energy level splitting ($qV$) and the operating temperature characterize the emission. In this way, the electrical power of a PV exposed to the outside at a corresponding operating temperature, defined as the steady state temperature or operating temperature, is self-consistently determined by obtaining the solution of Equation (6) with $P_{net,cool} = 0$ for a solar cell operating at the maximum power point ($V=V_{mp}$).

In the current work we consider a conventional radiative cooler (glass slab of thickness 3.2 mm and material parameters given by Palik[18]) with an absorptivity/emissivity, $\varepsilon(\lambda,\theta)$, that is calculated by performing full-wave electromagnetic simulations for wavelengths from 4 up to 33 μm with a 5° angular resolution using the commercially available software CST Microwave Studio. Moreover, we determine the absorptivity/emissivity of the cell $\varepsilon_{Si}(\lambda)$ by the data deduced from Fig. 1b from Refs[13,17].

## 4. Physical mechanisms, requirements and potential for cooling radiatively photovoltaics

To evaluate/validate our approach initially we calculated the efficiency and the open circuit voltage changes with respect to the operating temperature variations assuming the aforementioned theoretical model for the mono-crystalline silicon-based PV. We compared our calculated power-temperature and voltage-temperature coefficients (i.e., the slopes of the $P_{ele,max}$-$T$, $V_{OC}$-$T$ curves) with those of commercial PVs measured and provided by the manufacturers, and we verified very good agreement (the slopes of the $P_{ele,max}$-$T$, $V_{OC}$-$T$ curves are normalized at % compared to a PV operating at Standard Test Conditions (STC) (i.e., 1000 W/m² irradiance, AM 1.5G, $T_{cell}$=298.15K)). In particular, we calculated, for our theoretical PV (with silicon data obtained from Refs[24,25], see Section 3), a constant power-temperature coefficient equal to 0.293%/K and a voltage-temperature coefficient equal to 0.244%/K, values which are typical in commercial mono-crystalline silicon-based PV systems[33,34]. In this way, we confirm that the efficiency changes with temperature are linear. Moreover, our calculations showed that the efficiency changes are mainly controlled by the voltage changes with temperature; this is a consequence of the increased carrier concentrations at elevated temperatures and hence increased nonradiative (Auger) recombination[2], further supporting the validity of our assumptions discussed in Section 3. Indicatively, assuming only radiative recombination, the power-temperature and voltage-temperature coefficients are reduced almost down to the half (0.168%/K, 0.117%/K), implying a significant underestimation of the efficiency increase (almost half) related to the temperature reduction that could be provided by a radiative cooler. The excellent agreement of our theoretical calculations with the experimental data provided by the PV manufacturers allowed us to continue with the examination of the radiative cooling impact on the efficiency assuming a PV which operates outdoors.

After determining the dependence of the PV efficiency on temperature, we need next to relate the temperature with the power that either cools or heats the solar cell. Such a relation provides an effective way to evaluate the cooling provided by different cooling approaches and to compare different approaches/coolers. The impact of a certain amount of cooling power or vice versa of heating power on the operating temperature depends upon the slope of the total cooling power – temperature curve (for different weather conditions), where the total cooling power [$P_{cool}(T) = P_{rad,cooler}(T) - P_{atm}(T_{amb}) + P_{cond+conv}(T_{amb}, T)$] varies exponentially with temperature owing to the exponential dependence of $P_{rad,cooler}(T)$ with $T$ (see Equation S2). The weather conditions are included in the model by considering different combined conduction-convection nonradiative heat transfer coefficients, $h_c$, in the $P_{cond+conv}$ term of Eq. (6) and different $T_{amb}$ in the $P_{atm}$ term. For example, in windy conditions $h_c$ increases. In Fig. 2, we present the total cooling power for $T_{amb}$=300 K with respect to the cooler's temperature $T$ for two types of coolers, (i) the conventional thermal emitter used in commercial silicon PVs consisting of flat fused quartz (solid lines) and (ii) assuming a theoretical ideal thermal emitter, i.e., one exhibiting maximum emissivity along the entire thermal wavelength (4-33 μm) range for all angles of incidence (dashed lines).

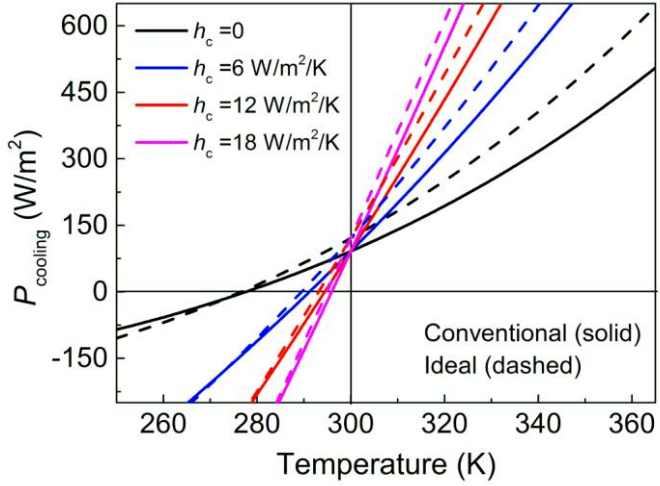

*Fig. 2: Total cooling power vs. cooler's temperature T for $T_{amb}$=300 K (vertical black line) and for different combined conduction-convection nonradiative heat transfer coefficients, $h_c$, for a conventional (flat fused quartz) thermal emitter (solid lines) and an ideal thermal emitter, i.e., exhibiting maximum emissivity along the entire thermal wavelength range (4-33 μm) for all angles of incidence (dashed lines).*

As it is confirmed in Fig. 2, the cooling power increases exponentially with temperature. However, especially for higher values of $h_c$, $P_{cool}(T)$ has a more linear-like response in the typical operating temperature range of PVs (310 K - 345 K). Thus, a certain amount of power (either extracted or added to the system) has a certain impact on temperature (reduction or increase) which depends upon the magnitude of the slope of the $P_{cooling}$-$T$ curve for each line/cooling condition. Accordingly, when we reflect the parasitic UV and sub-bandgap absorbed radiation in PVs, the impact of the reflected power on the operating temperature reduction depends upon the slope/derivative of the $P_{cooling}$-$T$ curve corresponding to each cooling condition (i.e., for each $h_c$ and for each cooler). Lower derivatives indicate higher temperature reduction for the same power reflected. In this way, we can also determine coefficients related to the cooling/extracted power needed for 1 K temperature reduction for each weather condition. Assuming realistic weather conditions (i.e., $T_{amb}$=300 K, $h_c$~12W/m$^2$), the $P_{cooling}$-$T$ coefficient (slope) equals to 17.9 W/m$^2$/K or 19.3 W/m$^2$/K for the conventional or the ideal thermal emitter respectively. The $P_{cooling}$-$T$ linear-like response further indicates that the power variations at the system play the most crucial role rather than the exact initial value of its heating power. This is not valid though for solar cells operating at lower operating temperatures, i.e., higher bandgap solar cells or including low-absorbing encapsulation layers, and assuming low-wind conditions, where the $P_{cooling}$-$T$ curve does not approach a linear-like response.

From Fig. 2, it is also clear that when we alter the cooling system, that is when we optimize the conventional thermal emitter towards the ideal, the cooling power increases and so does the slope, which is evident from the power difference between the solid and the dashed curves. Increased cooling power indicates an "in" and "out" power balance ($P_{net,cool}(V_{mp}, T)=0$) at lower temperatures. This justifies the increased interest in optimizing radiative-coolers over the recent years. However, passive radiative cooling (PRC) impact, that is the power difference between the conventional and the ideal thermal emitter, decreases as $h_c$ increases, as can be seen in Fig. 2 by comparing the solid lines versus the dashed curves. Interestingly though, PRC impact is still prominent even for high $h_c$ values (>12 W/m²/K). Moreover, due to the increased heating in a commercial c-Si PV which absorbs incident radiation almost ideally for a broad wavelength range within the absorption band of silicon (see Fig. 1b), where sun has its maximum intensity, and due to the high parasitic absorption at the various layers, the heating will be increased and the cell will operate at higher temperatures where the PRC impact is higher for all weather conditions.

We notice in Fig. 2 two crossing points of the lines concerning the different cooling conditions (different $h_c$), at $T=T_{amb}=300$ K, one for each of the two cooling systems. The meaning of these crossing points is that for $T=T_{amb}=300$ K the nonradiative heat transfer, i.e., due to convection (third term in right-hand side of Eq. (6)) is zero due to the nonradiative thermal equilibrium. The extra cooling power provided by each system is only due to the radiative heat transfer, which is still very high owing to the radiative access (through the atmospheric transparency window) to the universe with a much lower temperature (~3 K) than that of the atmosphere (~300 K). Indicatively, for $T=300$ K, the ideal thermal emitter provides a 124 W/m² cooling power compared to that of the conventional thermal emitter of 93 W/m². For lower cooler's temperatures than 300 K, the behavior in respect with $h_c$ inverses and the net cooling power starts again to decline until it obtains negative values (from 295 K - 276 K and below). However, since such low temperatures do not apply to photovoltaics, we are mainly interested in the regime $T>300$ K.

In Fig. 3 we present the impact of different radiative cooling approaches (among the ones discussed in the Introduction) on both the temperature reduction (Fig. 3a) and the efficiency enhancement (Fig. 3b) of the PV, calculated by employing the coupled thermal-electrical modeling proposed in this work (comparing to the same PV without the implemented radiative cooling approach). The cooling approaches presented include reflection of parasitic UV radiation (*UV* – black lines), reflection of the sub-bandgap radiation (*Sbg* – magenta lines), implementation of an ideal mid-IR thermal emitter (*Ideal* – green lines), and combinations of all the above. In the combined case of the reflection of both UV and sub-bandgap radiation and the additional implementation of the ideal thermal emitter (purple lines) the effect of changing the $T_{amb}$ (triangles) and the silicon thickness (*W*) (circles) is also demonstrated. The cases with climates with very weak winds or assuming protective windshields are found in the regime lower than $h_c=10.6$ W/m²/K which in Fig. 3a and b are at the left of the corresponding vertical lines. Cases with stronger winds are found in the right of the vertical $h_c=10.6$ W/m²/K lines.

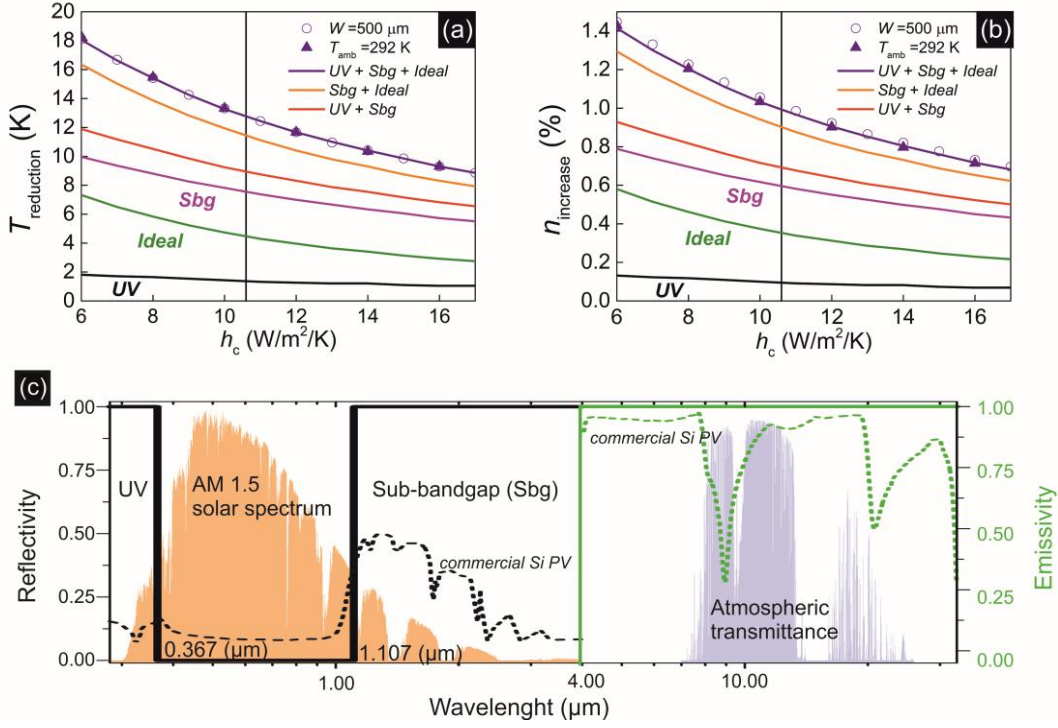

*Fig. 3: (a) PV temperature, T, reduction and (b) efficiency, η, increase associated to different radiative cooling approaches with respect to the combined conduction-convection nonradiative heat transfer coefficient $h_c$ (the reduction and increase are relative to the PV without any implemented cooling approach). Black lines correspond to the reflection of parasitic UV assuming an IQE=1, magenta lines correspond to the reflection of the sub-bandgap radiation, green lines correspond to the implementation of an ideal mid-IR thermal emitter, i.e., exhibits maximum emissivity along the entire thermal wavelength range (4-33 μm) for all angles of incidence, red lines correspond to the reflection of both UV and sub-bandgap radiation, orange lines correspond to the reflection of sub-bandgap radiation and the additional implementation of the ideal thermal emitter. Purple lines correspond to the reflection of both UV and sub-bandgap radiation and the additional implementation of the ideal thermal emitter. Triangles show the effect of the last approach for different $T_{amb}$ (i.e. 292 K instead of 300 K) and circles for different silicon thickness (W) (i.e. 500 μm instead of 250 μm). Cases at the left of the vertical line correspond to climates with very weak winds or assuming protective windshields. (c) Optimum reflectivity and emissivity spectrum (solid lines) for cooling radiatively a commercial crystalline-silicon PV in comparison with PV's reflectivity and emissivity (dashed lines).*

For calculating the results related to the reflection of the UV radiation, we introduced an algorithm in the theoretical modeling. This algorithm gradually generates a total reflection (that equals unity) from wavelength equal to 0.28 μm (highest thermalization losses) till the optimum wavelength, with a wavelength step of 0.0003 μm. In this way, we examine the relative contribution of the UV spectrum on the efficiency of a PV, considering also that the thermalization process at that regime is quite prominent as the excess energy of photons relative to the bandgap of the semiconductor is high, in addition to the high parasitic absorption from EVA. We found that the reflection of the UV radiation led to an increase (by ~0.1%) rather than a decrease of the PV efficiency, despite the reflection of potential currents. In other words, the negative effects of EVA absorption and thermalization losses seem to overcompensate the positive effect of the additional potential currents generated by the UV. For all weather conditions the optimum reflection wavelength range was found from 0.28 μm to 0.367 μm (EVA parasitic absorption nearly 0.8). As a result, as concluded by our analysis and can be seen in Fig. 3, the cost of the existing techniques for screening harmful[35] UV radiation and surface passivation techniques[36] (due to defects at the front surface of the cell acting as recombination traps) could be reduced considerably as well as the aging rate[37] since both an efficiency increase and a temperature reduction can occur despite reflecting incident UV radiation within the absorption band of silicon.

From Fig. 3a and 3b, it is also clear that the most efficient cooling approach in crystalline silicon-based PVs is the reflection of sub-bandgap radiation, due to the relatively high parasitic absorption from the PV module at that regime. We note that the cooling impact from the reflection of the sub-bandgap radiation could be more prominent if combined with the reflection of UV radiation. Moreover, as we see in Fig. 3a, an ideal thermal emitter provides also a significant temperature reduction (see also Fig. 2). Combining all radiative cooling approaches, a crystalline silicon-based PV can reach ideally a temperature reduction of up to 18 K, corresponding to an ~1.42% overall efficiency increase (compared to a PV where no any cooling approach has been applied), and up to 12.7 K corresponding to an ~1.0% overall efficiency increase assuming more realistic operating conditions ($h_c \geq 10.6$ W/m$^2$/K). In the solar cell industry, such an improvement is expected to lead to an increased lifetime[5] of the solar cell array, more than doubled, and an increased profit.

Operating temperature reduction remains almost the same if assuming an 8 K lower ambient temperature as seen in Fig. 3a. Interestingly, although for $T_{amb}$=292 K the PV operates at ~7 K lower temperature than for $T_{amb}$=300 K, the temperature reduction offered by the cooler does not decline. These results suggest that the radiative cooling strategy could be effectively utilized no matter the climate. Moreover, assuming a much thicker silicon layer in the PV, i.e. of $W$=500 μm, the operating temperature reduction remained almost the same. However, the efficiency increase for the case of a PV with $W$=500 μm is slightly higher (up to 0.032%), compared to that of $W$=250 μm, due to the alteration of the voltage- (0.253%/K) and power-temperature (0.303%/K) coefficients of the device arising from the nonradiative recombination rate dependence on $W$ (see Equation (4)). Consequently, we conclude that the radiative cooling of PVs is a very robust strategy to increase their efficiency, in respect with the varying operating conditions and the various characteristics that are met in commercial PVs.

Treating the PV as a quantum device instead of a solar absorber, i.e. taking into account also the electrical power generated by the incoming radiation besides the heating effect of that radiation (see Eq. (7)), we noticed up to ~1.1K smaller temperature reduction (corresponding to a ~0.32% lower power output) when an ideal thermal emitter was applied instead of the conventional thermal emitter. The reason was the lower amount of solar heating power, due to the electrical power output from the PV, which led the device to operate at lower temperatures, where the PRC impact is lower. Thus, since the optimization of the thermal emitter affects the nonradiative recombination rate only through the temperature, the efficiency increase in a PV due to the temperature reduction provided by the thermal emitter can be described with an error up to 1.1K by employing the PRC modeling (treating the PV as a solar absorber) and the power-temperature and voltage-temperature coefficients of the PV manufacturers. On the other hand, optimizing a radiative cooler unavoidably results to changes in the optical response of the system; for example, it may affect greatly the transparency from the top surface of the PV and hence, the $J_{SC}$ and the efficiency. Therefore, as it will be shown and in the next section, when utilizing a radiative cooler in a PV device, it is also necessary to weigh the interplay between the requirements for transparency in the optical spectrum and the enhanced, broadband thermal emission in mid-IR, and the way that they affect the cooler's reliability and fabrication cost.

## 5. Towards realistic implementation

In this last section, we apply the theory presented in the previous sections in the case of realistic photonic coolers proposed in the literature to evaluate and examine how far the current realistic implementations have come towards the maximum potential of the radiative thermal management in commercial solar cells, i.e. how close they are to the ideal implementation of all the cooling approaches shown in Fig. 3. Several studies[10,11,12,13,14] utilizing photonic radiative coolers for solar cells have emerged over recent years. In our study we pick two of them to highlight and distinguish the cooling gain that arises in one case mostly by the photon management at the optical regime (Wei Li et al.[13] in 2017) and in the other case at the mid-IR (Linxiao Zhu et al.[10] in 2015). Zhu et al.[10] exploited a 2D photonic crystal (PC) on top of a solar absorber (with a structure that emulated the behavior of a real silicon solar cell) which consisted of periodically placed air holes (~10 μm depth, ~6 μm periodicity) of non-vertical sidewalls in silica (see bottom structure of Fig. 4a). The nonvertical sidewalls of the holes resulted in a gradual refractive index change which provided effective impedance matching between silica and air over a broad range of thermal wavelengths (see Fig. 4c) that persisted even for larger angles of incidence (see Fig. 4d). Moreover, this visibly transparent thermal black-body led to the increase of the absorbed solar power in silicon due mainly to the enhanced transparency from the top surface (see Fig. 4b). Later, Wei Li et al.[13] proposed a 1D photonic crystal consisting of 45 alternate $Al_2O_3$, SiN, $SiO_2$, $TiO_2$ thin-film layers that could be implemented as a retrofit to current photovoltaic modules (see top structure of Fig. 4a). This photonic coating layer was designed to be placed on top of a PV and simultaneously reflect part of the solar spectrum that does not contribute to the photocurrent, i.e., the UV, sub-bandgap parasitic absorption, and further enhance the beneficial optical absorption (see Fig. 4b) and the thermal radiation in the mid-IR (see Fig. 4c, 4d).

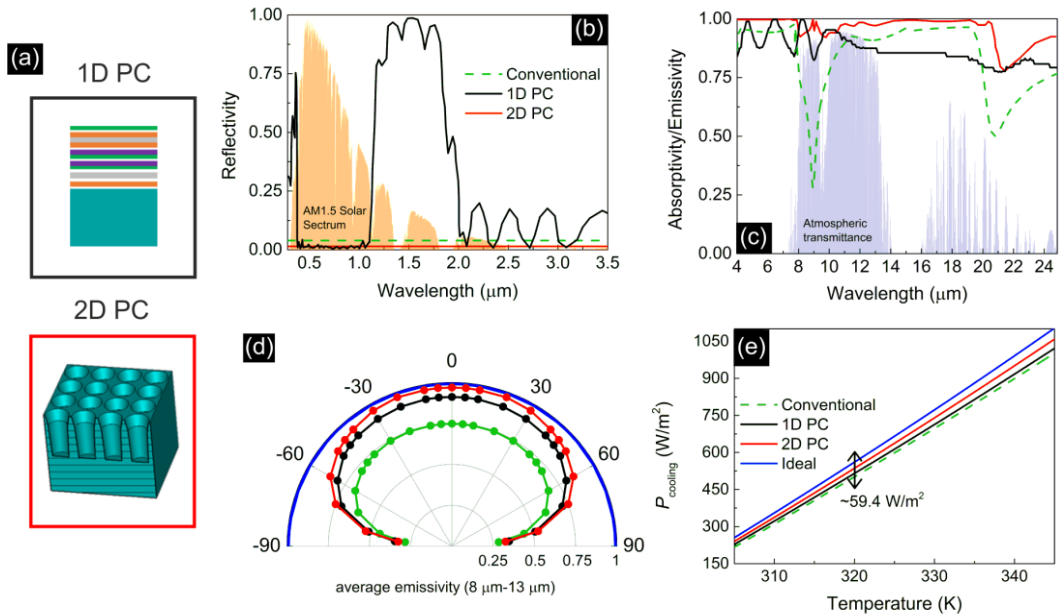

*Fig. 4: (a) Illustrations of a 1D photonic crystal consisting of alternate $Al_2O_3$, SiN, $SiO_2$, $TiO_2$ thin-film layers (top structure, in a black rectangle) and a 2D photonic crystal of non-vertical sidewalls in silica (bottom structure, in a red rectangle). (b) Reflectivity spectra of the 1D (black line) and 2D (red line) photonic crystals in comparison with the conventional case (flat fused quartz - green line) and (c) their emissivity spectra over the thermal wavelength range in mid-IR. Data are extracted from Refs.[10,13,18]. (d) Average emissivity between 8 and*

*13 μm (the atmospheric transparency window) plotted as a function of polar angle of incidence, for the 1D (black line) and 2D (red line) photonic crystal in comparison with the conventional (green line) and the ideal case, i.e., an overall ideal photonic cooler and not just an ideal thermal emitter (blue line). (e) Total cooling power ($P_{cool}$ (T) = [$P_{rad,cooler}$ (T) - $P_{atm}$ ($T_{amb}$) + $P_{cond+conv}$ ($T_{amb}$, T)]) vs. cooler's temperature T for each case for $T_{amb}$=298 K and for a combined conduction-convection nonradiative heat transfer coefficient equal to 13.7 W/$m^2$/K.*

The total cooling power of the above-mentioned 2D and 1D PC photonic coolers [$P_{cool}$ (T) = $P_{rad,cooler}$ (T) - $P_{atm}$ ($T_{amb}$) + $P_{cond+conv}$ ($T_{amb}$, T)] is a function of the reflectivity, absorptivity/emissivity integrals (see Appendix). Assuming $T_{amb}$=298 K and a nonradiative heat transfer coefficient equal to 13.7 W/$m^2$/K, to mimic typical outdoor conditions, the total cooling power (see Fig. 4e) for the 1D, 2D PC radiative coolers at T=320 K (typical operating temperature of commercial PVs) is calculated equal to ~514.6 W/$m^2$ and ~535.3 W/$m^2$ (ideally: ~560.2 W/$m^2$) respectively, whereas for the conventional PV it is 500.7 W/$m^2$. The 1D PC exploited a lower cooling power due to the lower emissivity than the 2D PC at the thermal mid-IR wavelength range (see Fig. 4c, 4d). On the other hand, the 1D PC provides a direct heat extraction, calculated as ~92 W/$m^2$ out of ~150 W/$m^2$, through the reflection of the parasitic UV and sub-bandgap solar absorption in optical (see Fig. 4b). Additionally, in Fig. 4e, it is clearly shown that for realistic operating conditions, the dependence of the cooling power versus temperature is linear. As such, a constant cooling-power-temperature coefficient for each case (19.6, 19.9, 20.5, 21.3 W/$m^2$/K for the conventional, 1D PC, 2D PC, and ideal case respectively) can be extracted. For instance, using the cooling-power-temperature coefficient for the 1D PC case, the temperature reduction is expected to equal ~4.7 K for the ~92 W/$m^2$ direct heat extraction through the reflection of the UV and the sub-bandgap radiation.

The roles of the 1D, 2D PC as radiative coolers in silicon PV modules were then investigated through current-voltage (J–V) calculations (see Equations (1), (3), (4)), and are shown in Fig. 5. In particular, we present the recombination current density (first two terms in the r.h.s, of Equation. (1)), Fig. 5a, and the output current density, Fig. 5b, for an operating temperature equal to the steady-state temperature arising by setting $P_{net,cool}$(V, T)=0 in Equation (6), assuming $T_{amb}$=298 K and a nonradiative heat transfer coefficient equal to 13.7 W/$m^2$/K to mimic typical outdoor conditions. The corresponding output electrical power and the steady-state temperature are presented in Fig. 5c and Fig. 5d respectively. Notice that as the electrical power increases the temperature drops due to heating reduction. The lowest operating temperature occurs at the maximum power point of the PV. The results are also compared with the ones that could be achieved ideally, that is assuming optimum performance in mid-IR and optimum reflection in optical, i.e., assuming an overall ideal photonic cooler and not just an ideal thermal emitter.

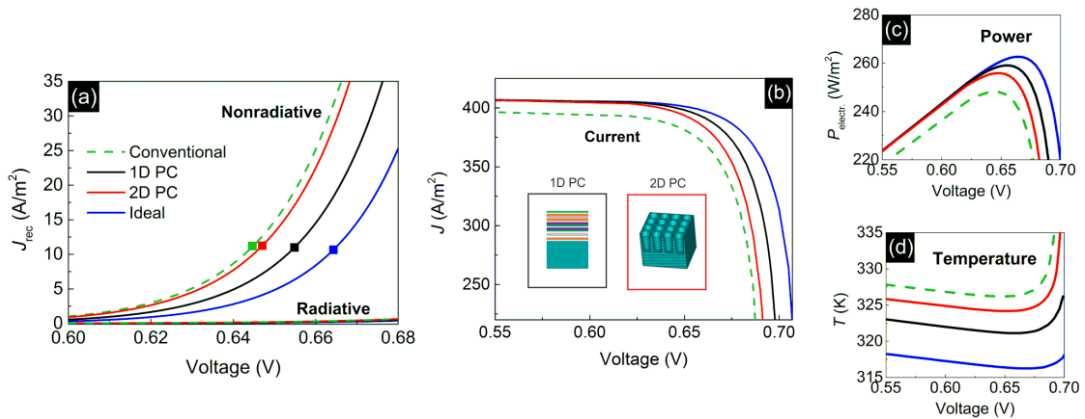

*Fig. 5: (a) Nonradiative and radiative recombination current density ($J_{rec}$) vs. applied voltage and the (b) current density vs. applied voltage of two PVs that incorporate a 1D photonic crystal (see top structure of Fig. 4a) and a 2D photonic crystal (see bottom structure of Fig. 4a) in comparison with the conventional PV (green line) and the ideal case assuming an overall ideal photonic cooler and not just an ideal thermal emitter (blue line). The squares in (a) denote the nonradiative (Auger) recombination current density at the maximum operating power point, $V_{mp}$, of the solar cell. (c) Electrical power output and the (d) operating temperature of the device vs. applied voltage for each case. Both (c) and (d) are calculated for the steady-state $P_{net,cool}(V, T)=0$.*

In Fig. 5b we observe that the useful current density increases for the ideal case and the photonic crystal radiative coolers cases in the voltage range of 0.55-0.71 V. This is also related to the enhanced short-circuit current density as calculated from Equation (2) which is increased from 396.4 A/m$^2$ to 406.9 A/m$^2$. (The short-circuit current density is equal to the current density at voltage=0 V, not shown in Fig. 5b). This ~2.6% short-circuit current density increase is due to the increased top surface transmissivity in the wavelength range within the absorptive band of silicon, provided from both photonic structures relative to the conventional flat glass. Moreover, despite the increased solar absorption (and hence the associated increased thermalization losses that resulted to higher heat dissipation in the structure), at the steady-state ($P_{net,cool}(V, T)=0$) the cooling properties of the 1D and 2D PCs result in operating temperature reduction compared to the conventional photovoltaic module. In particular, as seen in Fig. 5d, the temperature reduction at the maximum operating point of the PV ($V$s marked with squares in Fig. 5a) is equal to 5.1K for the 1D PC and 1.4K for the 2D (ideally: ~10K). This results in the increase of the open-circuit voltage, $V_{OC}$, by 1.4% and 0.42% respectively (ideally: ~2.6%).

The reason behind the $V_{OC}$ enhancement in all cases is the reduced nonradiative (Auger) recombination arising by the temperature reduction provided by the coolers. More specifically, as shown in Fig. 5a, the nonradiative recombination current density decreases when we optimize the cooling capability of the PV by employing the 2D, 1D PCs and the ideal photonic cooler. The reduced operating temperature provided by the coolers results to increase of the maximum power point voltages $V_{mp}$ (extracted from Fig. 5c) as we optimize the cooler from the conventional to the ideal case (see Fig. 5a). Moreover, one can see that the impact of the radiative recombination current density is much lower than the non-radiative in all cases. Results shown in Fig. 5a and b indicate improved dark current characteristics of the diode (the dark current is calculated from Equation (1) when the $J_{SC}=0$) that were achieved by the utilization of the photonic coolers since $V_{OC}$ increases as the saturation recombination current density decreases. Additionally, we note here, that with the increase of $J_{SC}$ the thermalization losses process increases too which slightly lowers $V_{mp}$ in all cases. However, the impact of this lowering on the efficiency was much smaller compared to the $J_{SC}$ contribution itself. Moreover, the fill factor *FF* is also improved by 0.03%, 0.3%, 0.6% for the 2D, 1D PCs and the ideal case respectively. Eventually, the conversion efficiency is further increased, due to the $J_{SC}$ increase and the $V_{OC}$ increase, and hence the improved dark characteristics of the diode. In particular, as also seen at the maximum power points of Fig. 5c, the conversion efficiency is increased by 3.1% in relative terms for the 2D PC case and by 4.3% for the 1D PC case, leading to higher overall efficiencies by 0.77% and 1.08% respectively with respect to the conventional PV; these values are quite close to the ideal case, calculated at ~+1.44%. Interestingly, aiming primarily for an enhanced thermal-emitter-cooler compared to an optical-reflector-cooler leads to an ~0.31% overall efficiency decrease.

## 6. Conclusions

In this work we examined and discussed the physical mechanisms and requirements for cooling radiatively commercial silicon PVs. This is done by employing a detailed electrical-thermal co-model which takes into account all the major processes affected by the temperature variation in a PV device. The accuracy and applicability of our model was tested and verified through comparison with experimental data provided by PV manufacturers, showing very good agreement. This confirmed the potential of the model to examine the dependence of the electrical properties of a PV device on temperature theoretically, for both indoor and outdoor conditions.

Employing our electrical-thermal model we found that the main reason of the efficiency decrease due to the heating in crystalline silicon-based PVs is the increased nonradiative recombination at elevated temperatures, which reduces the open-circuit-, and the maximum-point-voltage. Examining the relative potential of the different possible radiative cooling approaches in such PVs we found that the most efficient approach is the reflection of sub-bandgap radiation, due to the relatively high parasitic absorption from the PV module at that regime. Moreover, our study showed that the reflection of the UV radiation can also lead to decreased PV operating temperature and enhanced efficiency.

Finally, we found that the radiative cooling is a quite robust strategy to increase the PV efficiency, in respect to the varying operating conditions and the various characteristics of commercial PVs. Further improving the radiative cooling utilizing photonic structures can reduce the PV operating temperature up to ~10K (ideally) and hence enhance the efficiency up to ~5.8%, compared to PVs with conventional coolers (flat glass).

## Appendix

The steady-state temperature or the operating temperature of the cell of a photovoltaic (PV) module can be accurately described by treating the PV as a uniform device by using appropriate combined conduction-convection heat transfer coefficients. A thermal analysis for the PV can thus be performed by balancing the total power into and out the device following Planck's blackbody formalism and Kirchhoff's law, i.e., absorptivity equals emissivity, as is described in Section 3 of the main text.

Following Planck's formulation, the photon flux ($\Phi_{BB}$) and the spectral irradiance ($\varphi_{BB}$) of a blackbody at a temperature $T$ can be well accounted by:

$$\Phi_{BB}(T,\lambda) = (2\pi c/\lambda^4)/(e^{hc/\lambda k_B T} - 1) \quad (S1a)$$

$$\varphi_{BB}(T,\lambda) = (2hc^2/\lambda^5)/(e^{hc/\lambda k_B T} - 1) \quad (S1b)$$

where $h$ is Planck's constant, $k_B$ is Boltzmann's constant, $c$ is the speed of light. The power density (W/m²) radiated from a surface, in our case from the surface of the cooler, is then given by

$$P_{rad,cooler}(T) = \int d\Omega \cos\theta \int_0^\infty \varphi_{BB}(T,\lambda)\varepsilon(\lambda,\theta)d\lambda \quad (S2)$$

Here $\int d\Omega = \int_0^{\pi/2} d\theta \sin\theta \int_0^{2\pi} d\varphi$ is the angular integral over a hemisphere, and by using Kirchhoff's radiation law we replace the structure's absorptivity $\alpha(\lambda, \theta)$ by its emissivity $\varepsilon(\lambda, \theta)$. The term

$$P_{atm}(T_{amb}) = \int d\Omega \cos\theta \int_0^\infty \varphi_{BB}(\lambda, T_{amb}) a(\lambda, \theta) \varepsilon_{atm}(\lambda, \theta) d\lambda \qquad (S3)$$

describes the absorbed by the cooler thermal radiation emitted from the atmosphere, where the angle-dependent emissivity of the atmosphere is given by: $\varepsilon_{atm}(\lambda, \theta) = 1 - t(\lambda)^{1/\cos\theta}$, and $t(\lambda)$ is the atmospheric transmittance in the zenith direction. The term

$$P_{sun} = \int_0^\infty \alpha_{Si}(\lambda, \theta_{sun}) \Phi_{AM1.5G}(\lambda) \cos\theta_{sun} d\lambda \qquad (S4)$$

is the total solar absorption power density by the cell, where the solar illumination is represented by $\Phi_{AM1.5G}(\lambda)$, the AM1.5 spectrum[20] and $\alpha_{Si}(\lambda)$ is the cell's absorptivity. In Equation (S4) we assume that the structure is facing the sun at a fixed angle $\theta_{sun}$. Thus, the term $P_{sun}$ does not have an angular integral, and the silicon layer's absorptivity $\alpha_{Si}(\lambda, \theta_{sun})$ is represented by its value at $\theta_{sun}$. $P_{sun}$ either dissipates into heat or results to beneficial electrical power (calculated using the method of detailed balance by Shockley and Queisser[1] described in Section 3 of the main text) and emitted power by the cell:

$$P_{rad,cell}(T) = \int d\Omega \cos\theta \int_0^\infty \varphi(\lambda, T, qV_{mp}) \varepsilon_{Si}(\lambda) d\lambda \qquad (S5)$$

The power density radiated by the surface of the cell over a hemisphere, $P_{rad,cell}(T)$, is also known as the non-thermal radiation emitted by the solar cell due to the consequence of the bandgap of the semiconductor material[32]. Consequently, both the quasi-Fermi energy level splitting ($qV$), i.e. the difference in the quasi-Fermi levels of electrons and holes (the term "quasi" is due to the non-equilibrium (i.e. under-solar-illumination) steady state), and the operating temperature characterize the emission. Following Wurfel's generalized Planck law[32], the emitted spectral irradiance, $\varphi$, under the applied bias voltage $V$ (for $E - qV \gg k_B T$, where $E$ is the energy in eV), is given by:

$$\varphi(V, T, \lambda) = \varphi_{BB}(T, \lambda) e^{qV/k_B T} \qquad (S6)$$

In Equation (S5) we assume that the solar cell is operating at the maximum power point ($V=V_{mp}$). Finally, $\varepsilon_{Si}(\lambda) = \alpha_{Si}(\lambda, \theta_{sun})$ is the emissivity of the silicon layer that is assumed independent of polar angle $\theta$, even if the front surface of silicon is flat, because of its high refractive index that refracts the incident light very close to perpendicular inside the solar cell.

## Acknowledgments

This work was supported by the National Priorities Research Program grant No. NPRP9-383-1-083 from the Qatar National Research Fund (member of The Qatar Foundation).